\documentclass[preprint,12pt,3p]{elsarticle}

\usepackage{amssymb}
\newdefinition{rmk}{Remark}

\usepackage{multirow}
\usepackage{booktabs}
\usepackage{hyperref}
\usepackage{graphicx}

\usepackage{epstopdf} 

\usepackage{caption}
\DeclareGraphicsExtensions{.eps,.mps,.pdf,.jpg,.png}
\graphicspath{{figures/}{../figures/}}

\usepackage{fancyhdr}

\usepackage{amsmath}
\usepackage{amsfonts}

\usepackage{dsfont}

\usepackage{array}
\newcolumntype{C}[1]{>{\centering\let\newline\\\arraybackslash\hspace{0pt}}m{#1}}

\newcommand{\dd}{{\rm d}}
\newcommand{\kB}{k_{\mathrm{B}}}

\newcommand{\Tr}{\mathrm{Tr}}
\newcommand{\E}{\mathbb{E}}
\newcommand{\Var}{\mathrm{Var}}

\newcommand{\Lg}{\mathcal{L}}

\newcommand{\TT}{\mathsf{T}}

\renewcommand{\vec}[1]{{\mathbf #1}}

\newcommand{\bv}{{\vec{b}}}
\newcommand{\e}{{\vec{e}}}
\newcommand{\x}{{\vec{x}}}

\newcommand{\q}{{\vec{q}}}
\newcommand{\p}{{\vec{p}}}

\newcommand{\A}{{\vec{A}}}
\newcommand{\B}{{\vec{B}}}

\newcommand{\F}{{\vec{F}}}

\newcommand{\M}{{\vec{M}}}

\newcommand{\X}{{\vec{X}}}
\newcommand{\Y}{{\vec{Y}}}

\newcommand{\WB}{{\mathrm{\mathbf{W}}}}

\newcommand{\GammaB}{{\boldsymbol{\Gamma}}}
\newcommand{\sigmaB}{{\boldsymbol{\sigma}}}
\newcommand{\SigmaB}{{\boldsymbol{\Sigma}}}

\newtheorem{proposition}{Proposition}

\usepackage{mathrsfs}

\usepackage{soul}
\usepackage{color}

\usepackage[titletoc,toc,title]{appendix}

\usepackage{mathdots}

\begin{document}

\begin{frontmatter}

\title{Transient subtraction method: An alternative approach for variance reduction in dissipative particle dynamics}

\author[label1]{Xinyi Wu}
\address[label1]{School of Mathematical Sciences, Lancaster University, Bailrigg, Lancaster, LA1 4YF, United Kingdom}
\ead{x.wu38@lancaster.ac.uk}

\author[label2]{Xiaocheng Shang\corref{cor1}}
\address[label2]{School of Mathematics, University of Birmingham, Edgbaston, Birmingham, B15 2TT, United Kingdom}
\ead{x.shang.1@bham.ac.uk}

\cortext[cor1]{Corresponding author.}

\begin{abstract}
  The efficient and accurate computation of transport coefficients remains a numerical challenge in dissipative particle dynamics (DPD), which can require substantial sampling to control statistical uncertainty. We extend recent results obtained in Langevin dynamics to consider the transient subtraction method for reducing the variance of transport coefficient estimations in DPD. The system is initialised by perturbing equilibrium samples, and its relaxation towards equilibrium is monitored under the reference dynamics. A reference trajectory, initialised at equilibrium and driven by the same Brownian noise, provides a control variate for the transient response. We apply the transient subtraction method to the computation of the mobility and the shear viscosity in a Lennard-Jones fluid in DPD. The shear viscosity is obtained from the time-integrated response of a Fourier observable associated with the longitudinal velocity. Numerical comparisons with the naive transient estimator show that the transient subtraction method reduces the variance at various integration times, and the reduction is more pronounced with a higher friction coefficient. These results indicate that the transient subtraction method is able to improve the statistical precision of transport coefficient estimations in DPD and reduce the number of trajectories needed to attain a prescribed standard error.
\end{abstract}


\end{frontmatter}


\pagenumbering{arabic}

\section{Introduction}
\label{sec:Introduction}

Molecular dynamics provides a computational framework for investigating the properties of interacting particle systems from their microscopic equations of motion, with applications in chemical physics, materials science, and biophysics~\cite{Frenkel2001,Allen2017}. For isolated systems, Hamiltonian dynamics describes evolution at a constant total energy, whereas simulations at a given temperature employ a thermostat to sample the corresponding equilibrium distribution~\cite{Leimkuhler2015b}. A common choice is Langevin dynamics, in which frictional and stochastic forces represent interactions with a heat bath~\cite{Leimkuhler2015a,Shang2015,Lei2016a,Sachs2017,Shang2019,Leimkuhler2019a,Duong2021}. In its standard formulation, these forces act on individual particles and do not conserve the total momentum, which can affect the representation of hydrodynamic interactions~\cite{Duenweg1993}. Dissipative particle dynamics (DPD) was introduced to describe fluid behaviour at mesoscopic scales~\cite{Hoogerbrugge1992} and was subsequently formulated in terms of stochastic differential equations consistent with equilibrium statistical mechanics~\cite{Espanol1995}. In this coarse-grained interpretation, the particles represent portions of a fluid, reducing the number of explicitly resolved degrees of freedom~\cite{Espanol2017}. The DPD thermostat consists of pairwise dissipative and random forces that depend on relative particle positions, with the dissipative force also depending on relative velocities. The resulting dynamics conserves the linear momentum and respects the Galilean invariance, making DPD applicable to the study of hydrodynamic behaviour in complex fluids and soft matter. Although soft repulsive potentials are often used in coarse-grained models~\cite{Leimkuhler2015,Leimkuhler2016a}, the same thermostat can be combined with other conservative interactions, including Lennard-Jones potentials~\cite{Shang2017,Wu2025}.

Quantitative characterisation of transport in DPD requires estimators of transport coefficients, including mobility, diffusivity, and shear viscosity. Green--Kubo relations express these coefficients in terms of time integrals of equilibrium correlation functions~\cite{Green1954,Kubo1957}, whereas Einstein--Helfand relations involve the long-time growth rates of fluctuations in integrated currents~\cite{Helfand1960,Espanol2009}. For DPD, the fluxes entering these relations depend on the transported quantity and can contain contributions from dissipative and random forces. Recent studies have shown that Green--Kubo and Einstein--Helfand relations can be used to compute transport coefficients in DPD, provided that the expressions used account for the contributions of dissipative and random forces to the transport process~\cite{Malaspina2023,Malaspina2024}. In particular, the treatment of the random stress contribution affects the statistical accuracy of the Green--Kubo estimation~\cite{Malaspina2024}. The sampling requirements of the shear viscosity estimators at equilibrium have also been investigated for polymer solutions~\cite{Panoukidou2021}. Although these studies support the use of equilibrium methods, the computational effort required to attain a prescribed statistical precision depends on the estimator and the system parameters.

In nonequilibrium molecular dynamics, a sustained perturbation drives the system away from equilibrium, and a transport coefficient is then inferred from the resulting response in the linear response regime. The shear viscosity calculations in DPD have employed stress measurements under imposed shear and measurements of velocity profiles generated by external forces~\cite{Backer2005,Boromand2015}. Increasing the perturbation can improve the signal-to-noise ratio, although nonlinear contributions to the response restrict the perturbation magnitudes suitable for estimating linear response coefficients~\cite{Allen2017,Palmer1994}. At small perturbations, normalisation by the perturbation magnitude can instead amplify the statistical uncertainty in the estimated coefficient~\cite{Stoltz2022,Blassel2026}. Stochastic Norton dynamics provides a complementary formulation: a response observable is constrained to a prescribed value, and the forcing observable required to maintain this constraint is measured. This approach was developed for Langevin dynamics~\cite{Blassel2023} and subsequently extended to DPD~\cite{Wu2025}. Despite the different formulations, statistical uncertainty remains relevant to the computational cost of estimating transport coefficients.

Transient methods provide another route to computing transport coefficients by following the evolution of a system towards a stationary state~\cite{Blassel2026,Spacek2025}. Depending on the formulation, this evolution proceeds either from a perturbed initial distribution towards equilibrium or from equilibrium towards a nonequilibrium stationary state under external forcing. Examples of the former include the momentum impulse relaxation method for the viscosity~\cite{Arya2000} and the approach-to-equilibrium molecular dynamics method for the thermal conductivity~\cite{Lampin2013}. 

The statistical errors associated with transport coefficient estimators have motivated the development of variance reduction strategies~\cite{Wagner1997,Duncan2016,Stoltz2022,Blassel2026}. Importance sampling modifies the sampling distribution and compensates for this change through appropriate weights~\cite{Oettinger1994,Melchior1995}, while control variates modify an estimator using quantities with known expectations~\cite{Roussel2019,Pavliotis2022}. Another strategy employs synthetic forcings to extend the regime of approximately linear response, allowing larger perturbations to be used without increasing the nonlinear contribution to the same extent~\cite{Spacek2023}. These approaches offer different means of reducing the sampling effort required to estimate transport coefficients.

The transient subtraction method provides a further strategy based on the joint simulation of a perturbed system and a reference system, as introduced by Ciccotti and Jacucci~\cite{Ciccotti1975} and subsequently discussed in~\cite{Ciccotti1979}. Spacek et al. developed a transient subtraction method for stochastic dynamics with additive noise and applied it to Langevin dynamics~\cite{Spacek2025}. In this method, a trajectory with perturbed initial conditions relaxes towards equilibrium while a reference trajectory, initialised at equilibrium, evolves under the same dynamics and Brownian noise. This synchronous coupling allows the reference response to serve as a control variate for the transient response. The resulting estimator can have a lower variance when the two responses remain sufficiently correlated as the perturbed system relaxes towards equilibrium.

In this article, we formulate the transient subtraction method for DPD and examine the effect of the position-dependent noise coefficient on the coupled estimator. For the additive-noise dynamics considered in~\cite{Spacek2025}, the stochastic terms cancel upon subtraction of the synchronously coupled equations. In DPD, the two noise coefficients are evaluated at the respective particle configurations and generally differ, even though the same Brownian increments drive both trajectories. The equation for their difference therefore retains a stochastic term. Under stated regularity assumptions on the drift, the noise coefficient, and the response observable, we derive a finite-time bound on the variance of the transient subtraction estimator. The analysis first bounds the mean squared distance between the coupled processes in terms of their initial separation and then uses this estimate to control the variance of the time-integrated response difference.

We investigate the method numerically for the computation of the mobility and the shear viscosity in a Lennard-Jones fluid in DPD. For the mobility, the initial momenta are perturbed according to the colour drifts. For the shear viscosity, we use a sinusoidal perturbation of the initial momenta and integrate the response of the corresponding Fourier observable. The resulting linear response coefficient is related to the shear viscosity through the expression derived in~\cite{Wu2025}. We compare the transient subtraction estimator with the naive transient estimator for different friction coefficients, integration times, and the number of trajectories. The numerical results indicate lower estimator variances over the integration intervals examined, and larger variance reductions with the higher friction coefficients tested.

The rest of the article is organised as follows. Section~\ref{sec:Formulation of DPD} specifies the reference DPD dynamics, the equilibrium distribution at fixed total momentum, and the linear response framework. In Section~\ref{sec:Transient subtraction method}, we present transient estimators and the finite-time variance analysis. A variety of numerical experiments are reported in Section~\ref{sec:Numerical experiments} to compare the performance of the naive transient and transient subtraction estimators for the computation of the mobility and the shear viscosity. Our findings are summarised in Section~\ref{sec:Conclusions} with further possible extensions.

\section{Formulation of DPD}
\label{sec:Formulation of DPD}

We first specify the equilibrium DPD dynamics that governs the evolution of both the perturbed and reference trajectories in the transient subtraction method. We then introduce the equilibrium distribution and the response observables used to express transport coefficients. Consider $N$ interacting particles in a $d$-dimension periodic box of side length $L$. Their positions and momenta are denoted by $\q \in \mathcal{D}= (L\mathbb{T})^{dN}$ and $\p \in \mathbb{R}^{dN}$, respectively, where $\mathbb{T}=\mathbb{R}/\mathbb{Z}$ is the one-dimensional torus. The set of all positions and momenta is termed the phase space, denoted by $\mathcal{E}=\mathcal{D}\times\mathbb{R}^{dN}$, with the state $\X=(\q,\p)\in\mathcal{E}$. We rewrite the equations of motion of the DPD system in~\cite{Wu2025} on $\mathcal{E}$ (known as the ``reference dynamics'') as follows:
\begin{equation}\label{eq:DPD}
  \dd \X = \bv(\X) \, \dd t + \sigmaB(\X) \, \dd\WB \, ,
\end{equation}
where the drift vector and noise matrix are respectively given by
\begin{equation}
  \bv(\X)=
  \left(\begin{array}{c}
  \M^{-1}\p \\
  -\nabla U(\q) - \gamma\GammaB(\q)\M^{-1}\p
  \end{array}\right) \in \mathbb{R}^{2dN} \, , \quad \sigmaB(\X) = \left(\begin{array}{c}
  0 \\
  \sigma\SigmaB(\q)
  \end{array}\right) \in \mathbb{R}^{2dN \times K} \, ,
\end{equation}
where $\M\in\mathbb{R}^{dN\times dN}$ is the diagonal mass matrix and $U(\q)$ is the potential energy. Moreover, the conservative force is $-\nabla U$, where $\nabla$ represents the gradient operator. The parameters $\gamma$ and $\sigma$ determine the strengths of the dissipative and random forces, respectively. The process $\WB$ is a standard $K$-dimensional Brownian motion with $K=N(N-1)/2$, and its components are mutually independent and indexed by the particle pairs $(i,\, j)$ with $1 \leqslant i < j \leqslant N$. Moreover, the matrices $\GammaB(\q) \in \mathbb{R}^{dN \times dN}$  and $\SigmaB(\q) \in \mathbb{R}^{dN \times K}$ describe the pairwise dissipative and random interactions and satisfy the following relation:
\begin{equation}\label{eq:FDR Matrices}
  \GammaB = \SigmaB\SigmaB^{\TT} \, ,
\end{equation}
where the explicit formulation of the matrix $\SigmaB$ can be found in~\cite{Wu2025}. Moreover, the following relation needs to be satisfied
\begin{equation}\label{eq:FDR}
  \sigma^{2} = 2\gamma\kB T \, ,
\end{equation}
where $\kB$ is the Boltzmann constant and $T$ is the equilibrium temperature. We consider translation-invariant conservative interactions, for which the total momentum is conserved by the DPD dynamics. At a fixed total momentum $\boldsymbol{\pi}=(\pi^k)\in\mathbb{R}^d$, the canonical ensemble with the density is represented by
\begin{equation}\label{eq:rho_beta}
  \rho_{\beta}(\X) = {Z}^{-1} \exp(-\beta H(\q,\p)) \times \prod_{k=1}^{d}\delta \left[ \sum_{i=1}^{N} p^{k}_{i}- \pi^{k} \right] \, ,
\end{equation}
where $Z$ is a suitable normalising constant (i.e., the partition function), $\beta^{-1}=\kB T$, and $H(\X) = \p^{\TT}\M^{-1}\p/2 + U(\q)$ represents the system Hamiltonian. The delta functions impose the total momentum constraint. Together with~\eqref{eq:FDR Matrices}, the relation~\eqref{eq:FDR} ensures that this measure~\eqref{eq:rho_beta} is invariant under the reference dynamics. It is worth mentioning that Shardlow and Yan established geometric ergodicity for a one-dimensional DPD model under suitable assumptions on the interactions and particle density~\cite{Shardlow2006}. Their result gives uniqueness of the equilibrium measure on the corresponding constrained state space and exponential convergence towards equilibrium in the sense specified by their ergodicity theorem. Although this provides a rigorous basis for equilibration in that setting, its extension to more general DPD systems in higher dimensions is highly nontrivial. For the response formulation below, we use equilibrium initial conditions and assume that the relevant correlations are integrable in time.

For a sufficiently smooth observable $\phi$, the infinitesimal generator of the reference DPD system~\eqref{eq:DPD} acts as
\begin{equation}\label{eq:equilibrium generator}
  \Lg\phi = \M^{-1}\p \cdot \nabla_\q \phi - \nabla U(\q) \cdot \nabla_\p \phi - \gamma\GammaB(\q)\M^{-1}\p \cdot \nabla_\p \phi + \frac{\sigma^2}{2}\SigmaB(\q) \left[ \SigmaB(\q) \right]^{\TT}:\nabla^{2}_{\p} \phi \, , 
\end{equation}
where $\nabla^2_\p \phi$ is the Hessian with respect to the momentum variables and $\A:\B=\Tr\left(\A^{\TT}\B\right)$ denotes the Frobenius inner product. The associated Markov semigroup describes the evolution of expected observables through
\begin{equation}
  \left( e^{t\Lg}\phi \right)(\X) = \E \left[ \phi \left( \X_{t} \right) \mid \X_{0} = \X \right] \, .
\end{equation}
For any integrable observable with respect to the equilibrium measure, we define its equilibrium expectation as follows:
\begin{equation}
  \E_{\rho_{\beta}}[\phi(\X)] := \int_{\mathcal{E}} \phi(\X) \rho_{\beta}(\X) \, \dd \X \, ,
\end{equation}
where the expectation is taken over initial conditions sampled from the equilibrium distribution and over the Brownian noise driving the reference dynamics. We consider response observables in the space of square-integrable functions with zero equilibrium mean with respect to the density on the Hilbert space
\begin{equation}\label{eq:Hilbert space}
  L^{2}_{0}\left(\rho_{\beta}\right) = \left\{ \phi \in L^{2}\left(\rho_{\beta}\right) \mid \E_{\rho_{\beta}}[\phi] = 0 \right \} \, . 
\end{equation}
Let $R\in L^{2}_{0}(\rho_{\beta})$ be the response observable and $S \in L^{2}_{0} \left(\rho_{\beta}\right)$ the conjugate response observable associated with the chosen perturbation. The observable $S$ is determined by the perturbation, whose role in constructing the perturbed initial distribution is specified in the next section. Assuming that the equilibrium correlation is integrable in time, the associated linear response coefficient admits the Green--Kubo representation as follows:
\begin{equation}\label{eq:Green--Kubo}
  \zeta = \int_{0}^{\infty} \E_{\rho_{\beta}} \left[ R \left( \X_{t} \right) S \left( \X_{0} \right) \right] \, \dd t \, .
\end{equation}

\section{Transient subtraction method}
\label{sec:Transient subtraction method}

In this section, we introduce the transient dynamics and the synchronous coupling construction, and derive a finite-time variance bound for the resulting transient subtraction estimator.

\subsection{Transient dynamics}
\label{subsec:Transient dynamics}

In the transient approach, a small perturbation is locally applied to the initial distribution of the DPD system, and the subsequent relaxation towards equilibrium is monitored under the reference dynamics~\eqref{eq:DPD}. More precisely, the equations of motion of the transient DPD system are governed by
\begin{equation}\label{eq:transient equilibrium}
  \dd \X_{t}^{\eta} = \bv\left( \X_{t}^{\eta} \right) \, \dd t + \sigmaB\left( \X_{t}^{\eta} \right) \, \dd \WB_{t} \, , \quad \X_{0}^{\eta} \sim \widetilde{\rho}_{\eta} \, ,
\end{equation}
where $\eta\in\mathbb{R}^+$ denotes the perturbation magnitude at the initial condition, $\X_{t}^{\eta}$ presents the corresponding phase-space state, and the probability density $\widetilde{\rho}_\eta$ is defined by the first-order perturbation of the equilibrium density $\rho_{\beta}$~\eqref{eq:rho_beta}, with the conjugate response $S$ as follows:
\begin{equation}\label{eq:transient density}
  \widetilde{\rho}_{\eta} = \left( 1 + \eta S\right) \rho_{\beta} + O\left(\eta^2\right) \, .
\end{equation}
In practice, the initial condition is generated by applying a transformation to equilibrium samples. This transformation, introduced below, yields the prescribed perturbation to the relevant order in $\eta$. Then, for a given response observable $R \in L^{2}_{0} \left( \rho_{\beta} \right)$, recalling the definition of the transport coefficient based on the Green--Kubo representation~\eqref{eq:Green--Kubo}, the transport coefficient $\zeta$ can be alternatively expressed by
\begin{equation}\label{eq:transient equilibrium transport}
  \zeta = \lim \limits_{\eta \rightarrow 0} \frac{1}{\eta} \int_{0}^{+\infty} \E \left[ R\left(\X_{t}^{\eta}\right) \mid \X_{0}^{\eta} \sim \widetilde{\rho}_{\eta} \right] \, \dd t \, .
\end{equation}
The derivation of this transient response expression from the Green--Kubo relation is discussed in~\cite{Spacek2025}. Provided that the transient dynamics \eqref{eq:transient equilibrium} is ergodic, the estimator of the transport coefficient~\eqref{eq:transient equilibrium transport} can be approximated by the following discretised sum over the finite integration time $T = N_{\mathrm{sim}} \Delta t$ by $N_{\mathrm{run}}$ different trajectories:
\begin{equation}
  \widehat{\zeta}_{\eta,N_{\mathrm{run}},T}^{\mathrm{tran}} = \frac{\Delta t}{\eta N_{\mathrm{run}}} \sum\limits_{n=1}^{N_{\mathrm{run}}} \sum\limits_{m=1}^{N_\mathrm{sim}} R\left( \X_{m\Delta t}^{\eta,n} \right) \, .
\end{equation}

\subsection{Synchronous coupling}

A great deal of effort has been devoted to improving the efficiency of dynamical simulations~\cite{Oettinger1994,Melchior1995,Wagner1997}, as well as to reducing the variance and the bias in the computation of transport coefficients~\cite{Duncan2016,Stoltz2022,Pavliotis2022}. Coupling methods, including the reflection coupling~\cite{Lindvall1986} and the sticky coupling~\cite{Eberle2019,Darshan2024}, provide additional approaches in this context by introducing correlations between jointly simulated processes while preserving their individual dynamics. In particular, a recent development for Langevin dynamics, known as the transient subtraction method, was proposed in~\cite{Spacek2025}, which combines the ideas from the transient dynamics and the synchronous coupling in order to reduce statistical fluctuations in the targeted response. In this construction, two processes $\left( \X_{t}^{\eta},\Y_{t}^{0} \right)$ are evolved synchronously, and then the difference between their response observables is evaluated. More specifically, the reference process $\Y_{t}^{0}$ is initialised at equilibrium, while $\X_{t}^{\eta}$ starts from a perturbed initial condition and subsequently relaxes towards equilibrium. The response evaluated along $\Y_{t}^{0}$ serves as a control variate for the transient response. Driving both processes with identical realisations of the Wiener processes is intended to induce a strong positive correlation between their response observables, thereby reducing the variance of the response difference. Following the transient dynamics~\eqref{eq:transient equilibrium}, the dynamics of the coupling $\left( \X_{t}^{\eta},\Y_{t}^{0} \right)$ is synchronously evolved with the same vector of Wiener processes as follows:
\begin{subequations}\label{eq:coupling}
\begin{align}
  \dd \Y_{t}^{0} &= \bv\left( \Y_{t}^{0} \right) \, \dd t + \sigmaB\left( \Y_{t}^{0} \right) \, \dd \WB_t \, , \quad \Y_{0}^{0} \sim \rho_{\beta} \, , \label{eq:coupling equilibrium} \\
  \dd \X_{t}^{\eta} &= \bv\left( \X_{t}^{\eta} \right) \, \dd t + \sigmaB\left( \X_{t}^{\eta} \right) \, \dd \WB_t \, , \quad \X_{0}^{\eta} = \Phi_{\eta}\left( \Y_{0}^{0} \right) \, , \label{eq:coupling transient}
\end{align}
\end{subequations}
where the transformation function $\Phi_{\eta}$ maps equilibrium initial conditions to the perturbed initial distribution, chosen such that for any bounded measurable observable $\phi$, we have
\begin{equation}
  \int_{\mathcal{E}} \phi \widetilde{\rho}_{\eta} \, \dd \X = \int_{\mathcal{E}} \phi \circ \Phi_{\eta} \rho_{\beta} \, \dd \X \, .
\end{equation}
There are two types of transformations as follows:
\begin{subequations}
\begin{align}
  \Phi_{\eta}^{1}(\q,\p) &= \left( \q, \p + \eta\F(\q) \right) \, , \label{eq:map order 1}\\
  \Phi_{\eta}^{2}(\q,\p) &= \left( \q, \p + \eta\F(\q) - \left( \eta^{2}/2 \right) S(\q,\p) \F(\q) \right) \, , \label{eq:map order 2}
\end{align}
\end{subequations}
where $\F\in\mathbb{R}^{dN}$ is the position-dependent external force. The first-order transformation~\eqref{eq:map order 1} leads to the first-order perturbation in~\eqref{eq:transient density} (see more details in~\cite{Spacek2025}). For the given response observable $R$, the transport coefficient can be expressed in terms of the difference between the coupled responses as
\begin{equation}\label{eq:coupling transport}
  \zeta = \lim \limits_{\eta \rightarrow 0} \frac{1}{\eta} \int_{0}^{+\infty} \E \left[ R\left(\X_{t}^{\eta}\right) - R\left(\Y_{t}^{0}\right) \mid \Y_{0}^{0} \sim \rho_{\beta}, \X_{0}^{\eta} = \Phi_{\eta}\left(\Y_{0}^{0}\right) \right] \, \dd t \, .
\end{equation}
Assuming that the coupled dynamics~\eqref{eq:coupling} is ergodic, we approximate the ensemble expectations in~\eqref{eq:coupling transport} using averages computed from the simulated trajectory pairs. This gives the transient subtraction estimator:
\begin{equation}
  \widehat{\zeta}_{\eta,N_{\mathrm{run}},T}^{\mathrm{sub}} = \frac{\Delta t}{\eta N_{\mathrm{run}}} \sum\limits_{n=1}^{N_{\mathrm{run}}} \sum\limits_{m=1}^{N_\mathrm{sim}} \left[ R\left( \X_{m\Delta t}^{\eta,n} \right) - R\left( \Y_{m\Delta t}^{0,n} \right) \right] \, .
\end{equation}

\begin{proposition}
\label{prop1}
Suppose that the coupled processes in~\eqref{eq:coupling} are well defined on $[0,T]$ and $R$ is globally Lipschitz with constant $\|R\|^2_{\mathrm{Lip}}$. Assume that constants $B_{\bv},\, B_{\sigmaB} \in \mathbb{R}^{+}$ exist such that the following inequalities hold for all $\X_{t}^{\eta},\Y_{t}^{0} \in \mathcal{E}$,
\begin{subequations}
\begin{align}
  \langle \X_{t}^{\eta} - \Y_{t}^{0}, \bv\left(\X_{t}^{\eta}\right) - \bv\left(\Y_{t}^{0}\right) \rangle & \leqslant B_{\bv} \left| \X_{t}^{\eta} - \Y_{t}^{0} \right|^{2} \, , \label{eq:assumption b} \\
  \left\| \sigmaB\left(\X_{t}^{\eta}\right) - \sigmaB\left(\Y_{t}^{0}\right) \right\|^{2} & \leqslant B_{\sigmaB} \left| \X_{t}^{\eta} - \Y_{t}^{0} \right|^{2} \, , \label{eq:assumption sigma}
\end{align}
\end{subequations}
where $\langle \cdot \rangle$ is the inner product defined as $\langle \A,\B \rangle = \A^{\TT}\B$ and the Frobenius norm $\|\cdot\|$ is defined as $\|\A\|^{2}=\Tr\left(\A^{\TT}\A\right)$. Then, for any $t \geqslant 0$, we have
\begin{equation}\label{eq:lemma}
  \E \left[ \left| \X_{t}^{\eta} - \Y_{t}^{0} \right|^{2} \right] \leqslant e^{\left( 2B_{\bv} +B_{\sigmaB} \right)t} \E \left[ \left| \X_{0}^{\eta} - \Y_{0}^{0} \right|^{2} \right] \, ,
\end{equation}
and
\begin{equation}
  \Var \left( \widehat{\zeta}_{\eta,N_{\mathrm{run}},T}^{\mathrm{sub}} \right) \leqslant \frac{\|R\|^{2}_{\mathrm{Lip}}}{N_{\mathrm{run}}\eta^{2}} \E \left[ \left| \X_{0}^{\eta} - \Y_{0}^{0} \right|^{2} \right] \left( \int^{T}_{0} e^{\left( B_{\bv} + B_{\sigmaB}/2 \right)t} \, \dd t \right)^{2} \, .
    \label{eq: var upper}
\end{equation}
\end{proposition}
The proof of Proposition~\ref{prop1} is provided in~\ref{appendix:prop1}. Under the Lipschitz condition of Proposition~\ref{prop1}, the mean squared distance between the two synchronously coupled trajectories is controlled by their initial separation. Moreover, for the first-order transformation $\Phi_{\eta}^{1}$~\eqref{eq:map order 1}, we have
\begin{equation}
  \E \left[ \left| \X_{0}^{\eta} - \Y_{0}^{0} \right|^{2} \right] = \eta^{2}||\F||^{2} \, .
\end{equation}
Consequently, for any fixed integration time $T$, Proposition~\ref{prop1} yields a finite-time variance bound that is uniform in $\eta$. The condition on the drift vector $\bv$~\eqref{eq:assumption b} is one-sided because only the positive contribution of this inner product needs to be controlled in the estimate; a negative value can reduce the drift contribution to the evolution of the squared distance. In the DPD setting~\eqref{eq:DPD}, the noise matrix $\sigmaB$ contains the pairwise direction vector $\e_{ij} = \left(\q_{i}-\q_{j}\right)/r_{ij}$, which is the unit vector pointing from particle $j$ to particle $i$, with $r_{ij} = \left|\q_{ij}\right| = \left|\q_{i}-\q_{j}\right|$ being the distance between the two particles.  Note that $\e_{ij}$ is undefined at $r_{ij}=0$. Away from particle overlap, $\sigmaB$ is locally Lipschitz provided that the weight functions are locally Lipschitz. If the particle configurations remain in a region where all interparticle distances are bounded such that $r_{ij} \geqslant \theta > 0$, where $\theta$ is a positive constant, then the matrix $\SigmaB$ is locally Lipschitz in this region. This local interpretation is consistent with the numerical setting considered below, where the Lennard-Jones potential used in the simulations introduces strong repulsion at short separations and therefore suppresses configurations with very small distances.

\section{Numerical experiments}
\label{sec:Numerical experiments}

In this section, we compare the performance of the naive transient estimator and the transient subtraction estimator for the computation of transport coefficients in DPD, focusing on the mobility and the shear viscosity.

\subsection{Simulation details}
\label{subsec:Simulation details}

Although the traditional formulation of DPD employs a soft potential (e.g., in~\cite{Shang2020}), other conservative interactions have also been widely used in DPD simulations. In this article, we consider the Lennard-Jones system, which was also used in~\cite{Wu2025}, with the total potential energy as follows: 
\begin{equation}
  U(\mathbf{q}) = \sum\limits_{i=1}^{N-1}\sum\limits_{j=i+1}^{N} \varphi(r_{ij}) \, ,
\end{equation}
where $\varphi(r_{ij})$ represents the pair potential energy between the two particles, obtained by truncating and shifting the underlying Lennard-Jones potential $u$ at the cutoff radius $r_{\mathrm{c}}$ so that the potential and its first derivative are continuous at the cutoff:
\begin{equation}
\varphi\left(r_{ij}\right) =
  \begin{cases}
    \displaystyle u\left(r_{ij}\right) - u\left(r_{\mathrm{c}}\right) - u'\left(r_{\mathrm{c}}\right) \left(r_{ij} - r_{\mathrm{c}}\right) \, , & r_{ij} < r_{\mathrm{c}} \, ;\\
    \quad \quad \quad \quad \quad \quad 0 \, , & r_{ij} \geq r_{\mathrm{c}} \, ,
  \end{cases}
\end{equation}
where $u$ is given by
\begin{equation}
  u(r) = 4\varepsilon \left[ \left( \frac{r_{0}}{r} \right)^{12} - \left( \frac{r_{0}}{r} \right)^{6} \right] \, ,
\end{equation}
where $\varepsilon$ and $r_{0}$ are two constants that set the energy and length scales
of the beads, respectively, in reduced Lennard-Jones units.

Unless otherwise stated, we simulated $N=500$ identical particles in a three-dimensional cubic box (i.e., $d=3$) with periodic boundary conditions, using the ABOBA method~\cite{Shang2020,Wu2025}. We used the following parameter set: $\varepsilon=r_{0}=m_{i}=1$ (for defining reduced units), where $m_{i}$ is the mass of the particle $i$, and $k_{\mathrm{B}}=T=1$. The particle density was $\rho_{\rm d}=0.85$, and the cutoff radius was $r_{\mathrm{c}}=2.5$. The box length was therefore fixed by $L=\left(N/\rho_{\rm d}\right)^{1/3}$. The particles were initially placed on a uniformly spaced cubic lattice. The initial momentum components were sampled from independent, identically distributed normal distributions with mean zero and variance $k_{\mathrm{B}}T$, consistent with the Maxwell--Boltzmann distribution for unit particle mass. We considered two friction coefficients, $\gamma=4.5$ and $\gamma=40.5$, which have been widely used in tests of DPD algorithms. These values allow us to compare the estimators under weaker and stronger dissipative coupling. Cell lists~\cite{Verlet1967} were employed in all simulations to accelerate the identification of interacting particle pairs within the cutoff radius.

The system was equilibrated for 10 reduced time units before data collection to reduce the influence of the initial configuration. Unless otherwise stated, dynamical quantities were then evaluated over a further 10 reduced time units using a timestep of $\Delta t=0.001$. Results were averaged over 1000 runs to reduce the sampling error in the estimated response observables. We also tested the second-order transformation $\Phi_{\eta}^{2}$ in~\eqref{eq:map order 2}. The results from the second-order transformation were similar to the first-order transformation $\Phi_{\eta}^{1}$~\eqref{eq:map order 1}, therefore only the results of the first-order transformation are presented in this article. All simulations were performed using the Julia package \texttt{Molly}~\cite{Greener2024} on the \href{https://lancaster-hec.readthedocs.io/en/main/intro.html}{High End Computing (HEC) Cluster} at Lancaster University. The reference values for the mobility and the shear viscosity were taken from the simulation of the DPD-Norton dynamics reported in~\cite{Wu2025} and used as benchmarks for assessing the transient estimators.

\subsection{Numerical results}
\label{subsec:Numerical results} 

In this subsection, we separately introduce the observables used to compute the mobility and the shear viscosity and present the corresponding numerical results.

\subsubsection{Mobility}

The mobility is proportional to the diffusion coefficient, a fundamental transport coefficient in physics, and has been extensively studied in the context of Langevin dynamics~\cite{Evans2008,Wang2013a,Leimkuhler2013c,Pavliotis2022,Spacek2023}. The response observable $R$ and the conjugate response observable $S$ used in the mobility calculation are given by
\begin{equation}\label{eq:transient mobility response}
  R(\q,\p) = \F \cdot \M^{-1}\p \, , \quad S(\p) = \beta \F \cdot \M^{-1} \p \, ,
\end{equation}
where the external force $\F(\q) = \F = \left(\F_1,\F_2,\cdots,\F_N\right)^{\TT} \in \mathbb{R}^{3N}$ is a constant vector field and is normalised such that $|\F|=1$. Following~\cite{Wu2025}, we use the colour drifts that act in opposite directions on the two halves of the particle population. Taking the direction of the drift to be the $x$-direction gives
\begin{equation}
  \F_{i} = \frac{1}{\sqrt{N}} \left( (-1)^{i}, 0, 0 \right)^{\TT}, \quad i=1,2,\dots,N \, ,
\end{equation}
where we take $N$ to be even so that the colour drifts preserve the total momentum.

\begin{figure}[tbp]
  \centering
  \includegraphics[scale=0.36]{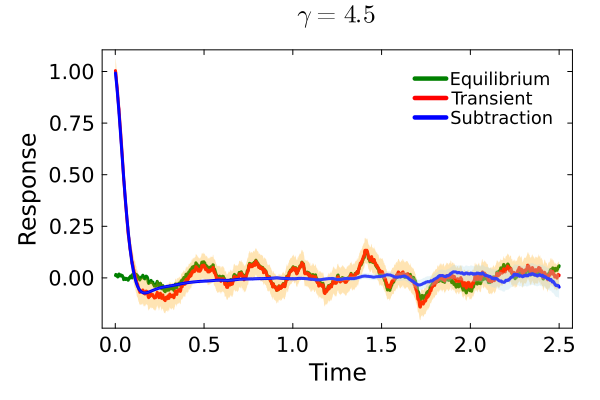}
  \includegraphics[scale=0.36]{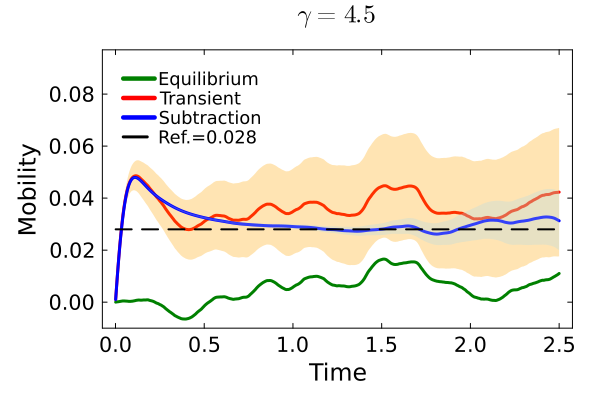}
  \includegraphics[scale=0.36]{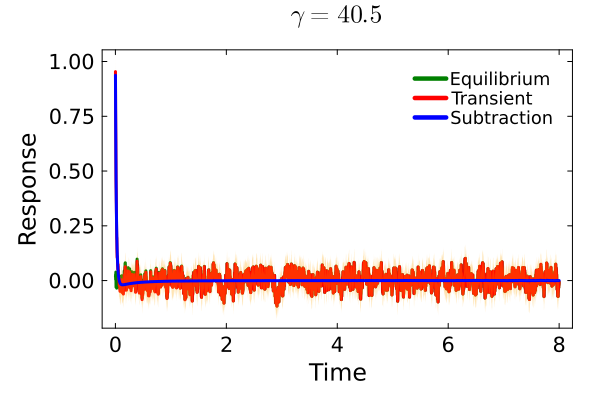}
  \includegraphics[scale=0.36]{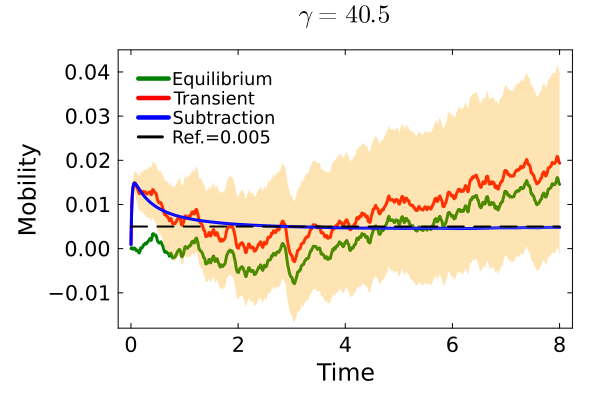}
  \caption{Trajectories of the response (left) and the associated estimator (right) for the mobility~\eqref{eq:transient mobility response} against time, with the error bars for the transient dynamics~\eqref{eq:coupling}, including the equilibrium process~\eqref{eq:coupling equilibrium}, the naive transient process~\eqref{eq:coupling transient}, and their subtraction, using the colour drifts with the strength $\eta=1$. Stepsize $\Delta t=0.001$ was used for the friction coefficients $\gamma=4.5$ (top) and $\gamma=40.5$ (bottom).}
  \label{fig:transient equilibrium mobility}
\end{figure}

In Figure~\ref{fig:transient equilibrium mobility}, we plot the trajectories of the response for the mobility against time on the left panels and the corresponding integration for the estimator~\eqref{eq:coupling transport} using the colour drifts with a fixed forcing strength $\eta=1$. The equilibrium process~\eqref{eq:coupling equilibrium}, the naive transient process~\eqref{eq:coupling transient}, and their response difference are shown in green, red, and blue, respectively. The system of the coupled equilibrium dynamics was simulated with the friction coefficients of $\gamma=4.5$ on the top panels and $\gamma=40.5$ on the bottom panels. Subtraction produces visibly smaller fluctuations and narrower uncertainty bands for both friction coefficients. More specifically, in the case of $\gamma=4.5$, the transient subtraction estimator approaches a plateau near the reference mobility at approximately $T\approx1$, which can be considered as the relaxation time, while fluctuations in the response difference become more apparent around $T\approx1.5$ showing the occurrence of the decoupling. This behaviour is consistent with a loss of correlation between the coupled responses after the initial relaxation. In contrast, for $\gamma=40.5$, the transient subtraction estimator stabilises near the reference value by approximately $T\approx3$, and the corresponding variance of the estimated response remains small up to $T=8$. The naive transient estimator continues to fluctuate and deviates from the reference value at later times within the simulated interval. This finite-sample behaviour indicates a larger sampling error, rather than establishing a failure of the naive estimator to converge. The comparison suggests that the transient subtraction method retains its statistical benefit for longer time in the high friction limit.

\begin{table}[tbp]
    \centering
    \setlength{\tabcolsep}{20pt}
    \begin{tabular}{cccc}
    \toprule
    \multicolumn{4}{c}{$\gamma=4.5$}\vspace{0.2cm}\\
    Time & Naive & Subtraction & Ratio  \\
    \midrule
    1 & $6.79\times10^{-2}$ & $1.64\times10^{-4}$ & $4.15\times10^{2}$ \\
    1.5 & $9.88\times10^{-2}$ & $1.27\times10^{-3}$ & 77.6 \\
    2 & 0.122 & $9.06\times10^{-3}$ & 13.4 \\ 
    \bottomrule  
    \end{tabular}
    
    \vspace{0.4cm} 
    
    \begin{tabular}{cccc}
    \toprule
    \multicolumn{4}{c}{$\gamma=40.5$}\vspace{0.2cm}\\
    Time & Naive & Subtraction & Ratio  \\
    \midrule
    3 &  $4.60\times10^{-2}$ & $1.10\times10^{-6}$ & $4.20\times10^{4}$ \\
    5 & $6.70\times10^{-2}$ & $3.45\times10^{-6}$ & $1.94\times10^{4}$ \\
    7 & $9.26\times10^{-2}$ & $1.07\times10^{-5}$ & $8.69\times10^{3}$ \\ 
    \bottomrule  
    \end{tabular}

    \caption{Comparisons of variances between the naive transient and transient subtraction estimators for the mobility in the transient dynamics~\eqref{eq:coupling} with friction coefficients of $\gamma=4.5$ and $\gamma=40.5$, corresponding to the results on the right panels in Figure~\ref{fig:transient equilibrium mobility}.}
    \label{tab:coupling mobility}
\end{table}

To quantify the variance reduction, Table~\ref{tab:coupling mobility} compares the variances of the naive transient and transient subtraction methods, using the results from the right panels in Figure~\ref{fig:transient equilibrium mobility} at selected times. Following~\cite{Spacek2025}, the variance ratio is defined as the variance of the naive transient estimator divided by that of the transient subtraction estimator, so that values greater than one indicate a benefit from the transient subtraction estimator. It can be seen from the data that for both friction coefficients considered, the variances of the mobility obtained from the transient subtraction estimator are substantially smaller than those obtained from the naive transient estimator, which confirms our observations in Figure~\ref{fig:transient equilibrium mobility}. Specifically, for $\gamma=4.5$, the variance reduction remains after fluctuations in the response difference have become apparent, although its magnitude diminishes with time. For $\gamma=40.5$, although the variance ratio decreases as time increases, it retains a substantial reduction beyond the initial relaxation. These results also suggest that extending the integration interval beyond the initial relaxation may introduce additional statistical noise without improving the estimation of the transient response, although the transient subtraction estimator retains a lower variance than the naive transient estimator over the times considered.

\begin{figure}[tbp]
  \centering
  \includegraphics[scale=0.36]{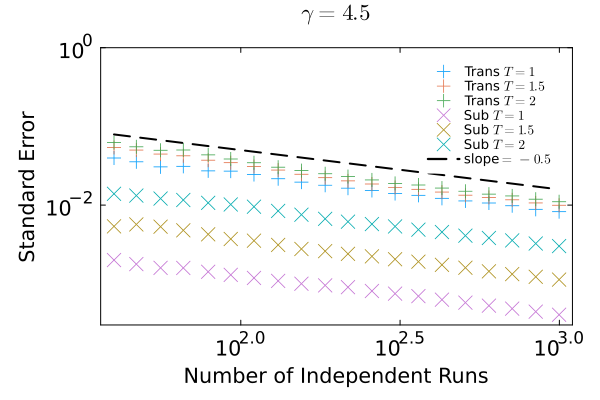}
  \includegraphics[scale=0.36]{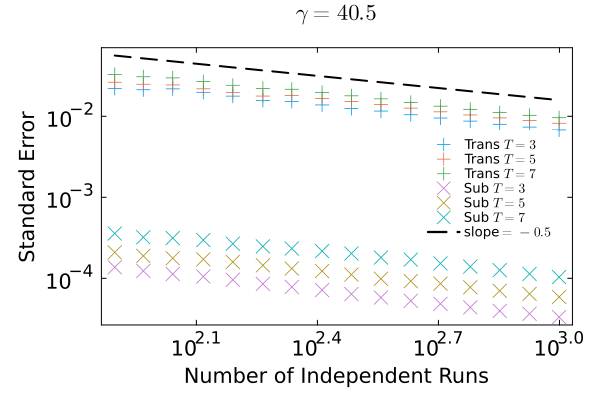}
  \caption{Standard errors of the response for the mobility~\eqref{eq:transient mobility response} against the number of trajectories for the naive transient method and the transient subtraction method, with friction coefficients of $\gamma=4.5$ (left) and $\gamma=40.5$ (right) by using the colour drifts with the strength of $\eta=1$, in a log–log scale. The expected scaling line is plotted in dashed lines in all cases}
  \label{fig:se mobility}
\end{figure}

To further investigate the statistical efficiency of the two estimators, Figure~\ref{fig:se mobility} examines the dependence of the standard error on the number of different trajectories at selected integration times on logarithmic scales. It can be clearly seen that the standard errors indeed scale as $N_{\mathrm{run}}^{-1/2}$ for both friction coefficients, while the transient subtraction method consistently produces smaller errors than the naive transient method. This scaling is consistent with the dependence on $N_{\mathrm{run}}$ in~\eqref{eq: var upper}. In the case of $\gamma=4.5$, the reduction in statistical error becomes less pronounced at later times, which is consistent with the decoupling of the two synchronously coupled trajectories observed in Figure~\ref{fig:transient equilibrium mobility}. In contrast, for $\gamma=40.5$, the transient subtraction method provides a substantially greater reduction in statistical error, even for long integration times. These observations are consistent with the variance ratios reported in Table~\ref{tab:coupling mobility}, indicating that the transient subtraction method is able to achieve a given level of statistical accuracy with fewer trajectories, particularly in the high friction limit.

\subsubsection{Shear viscosity}

The shear viscosity is another important transport coefficient considered in this article, which is commonly computed using nonequilibrium simulations with Lees--Edwards boundary conditions to generate a steady shear flow~\cite{Pagonabarraga1998,Shang2020}. An equilibrium alternative is to integrate the stress autocorrelation functions using a Green--Kubo relation~\cite{Chaudhri2010,Panoukidou2021}. Recently, a rigorous framework for how the shear viscosity of a three-dimensional DPD system can be obtained from the average longitudinal velocity profile was developed in~\cite{Wu2025}. Specifically, we consider an external force directed along the longitudinal $x$-direction, with an amplitude that depends on the transverse coordinate $y$. The relation used below follows from the linear response of the longitudinal velocity profile. In this regime, a spatially varying longitudinal force induces an average velocity profile with the same transverse modes, and the shear viscosity is given by
\begin{equation}\label{eq:shear viscosity computation}
  \nu = \frac{\rho_{\rm d} F_{1}}{U_{1}} \left( \frac{L}{2\pi} \right)^{2} \, ,
\end{equation}
where $F_{1}$ is the analytically known Fourier coefficient associated with the external force acting in the longitudinal $x$-direction, and $U_1$ is the Fourier coefficient of the average longitudinal velocity in the $y$-direction. For the numerical experiments tested in this section, the external force is defined by the sinusoidal force as follows:
\begin{equation}
  F(y) = \sin\left( \frac{2\pi y}{L} \right) \, ,
\end{equation}
and therefore the corresponding Fourier coefficient is $F_{1}=\mathrm{Im}(i/2)=1/2$. With synchronous coupling, the transient expression~\eqref{eq:coupling transient} gives the Fourier response $U_1$ with the associated response observable $R$ defined as
\begin{equation}\label{eq:transient viscosity response}
  R(\X) = \mathrm{Im}(R_{1}(\q,\p)) \, ,\quad R_{1}(\q,\p) = \frac{1}{mN} \sum\limits^{N}_{j=1} p^{x}_{j} \exp\left( \frac{2i\pi q^{y}_{j}}{L} \right) \, ,
\end{equation}
where $p^{x}_{j}$ is the momentum of the $j$-th particle in the $x$-direction and $q^{y}_{j}$ is the position of the $j$-th particle in the $y$-direction, and the conjugate response observable $S$ defined as
\begin{equation}
  S(\q,\p) = \beta \F\left(\q^{y}\right) \cdot \M^{-1}\p \, ,
\end{equation}
where $\q^{y}$ is the position vector in the $y$-direction.

\begin{figure}[tbp]
  \centering
  \includegraphics[scale=0.34]{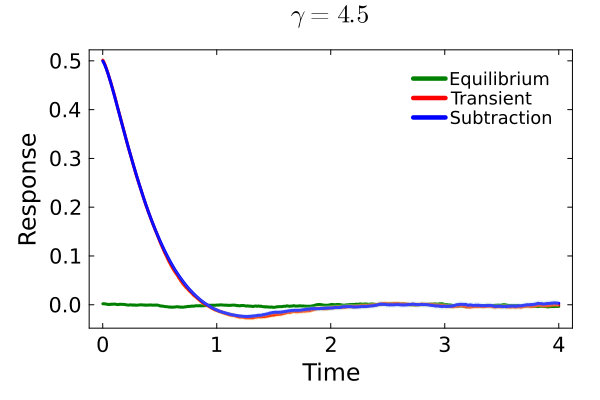}
  \includegraphics[scale=0.34]{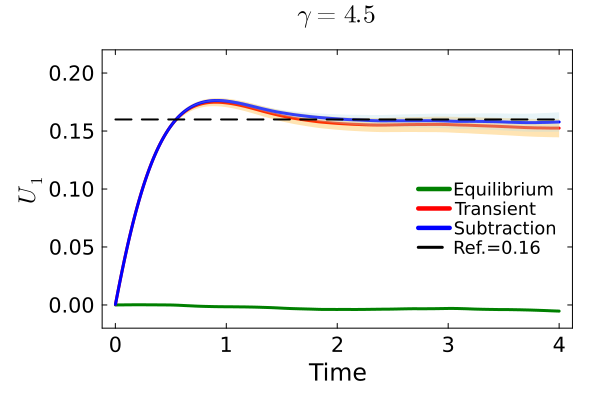}
  \includegraphics[scale=0.34]{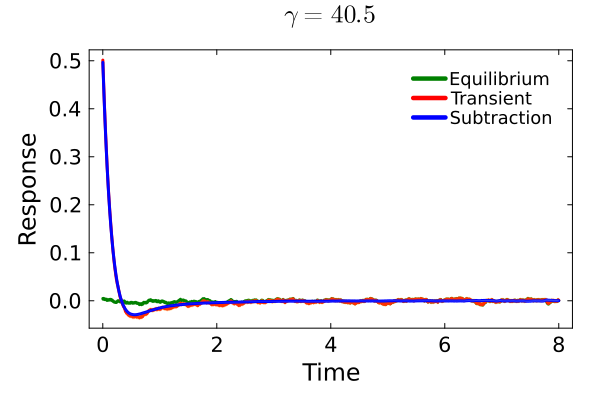}
  \includegraphics[scale=0.34]{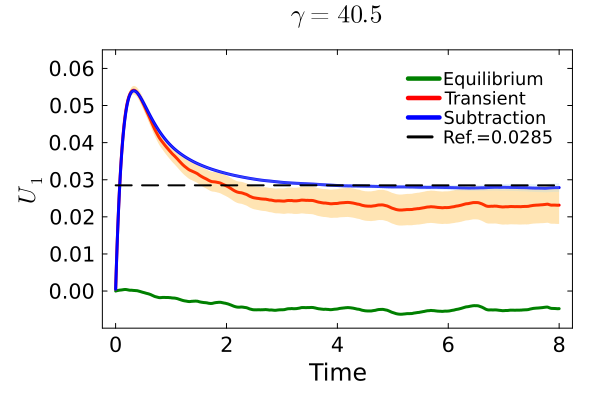}
  \caption{Trajectories of the response (left) and the associated estimator (right) for the Fourier coefficient $U_1$~\eqref{eq:transient viscosity response} against time, with the error bars for the transient dynamics~\eqref{eq:coupling}, including the equilibrium process~\eqref{eq:coupling equilibrium}, the naive transient process~\eqref{eq:coupling transient}, and their subtraction, using the sinusoidal force with the strength $\eta=0.4$. Stepsize $\Delta t=0.001$ was used for the friction coefficients of $\gamma=4.5$ and $\gamma=40.5$.}
  \label{fig:transient equilibrium viscosity}
\end{figure}

In Figure~\ref{fig:transient equilibrium viscosity}, the left panels display the time traces of the time-dependent response $R$ of the shear viscosity associated with~\eqref{eq:transient viscosity response}, while the right panels show the corresponding time integrals for the Fourier response $U_1$ used in the estimator~\eqref{eq:shear viscosity computation}, with friction coefficients of $\gamma=4.5$ and $\gamma=40.5$ in the same format as in Figure~\ref{fig:transient equilibrium mobility}. As shown in the plots, both the naive transient estimator and the transient subtraction estimator approach approximately stationary values over the simulated interval, but the transient subtraction estimator lies closer to the reference value at later times. In particular, for $\gamma=40.5$, the naive transient estimator remains visibly below the reference value, so the appearance of a plateau alone does not establish convergence to that value. The transient subtraction estimators settle near their plateaux by roughly $T\approx2$ for $\gamma=4.5$ and $T\approx4$ for $\gamma=40.5$, respectively, whose times characterise the observed relaxation of the integrated response. 

\begin{table}[tbp]
    \centering
    \setlength{\tabcolsep}{20pt}
    \begin{tabular}{ccccc}
    \toprule
    \multicolumn{4}{c}{$\gamma=4.5$}\vspace{0.2cm}\\
    Time & Naive & Subtraction & Ratio  \\
    \midrule
    2 & $7.65\times10^{-3}$ & $3.88\times10^{-3}$ & 1.97 \\
    2.5 & $9.63\times10^{-3}$ & $6.97\times10^{-3}$ & 1.38 \\
    3 & $1.15\times10^{-2}$ & $1.00\times10^{-2}$ & 1.15 \\ 
    \bottomrule  
    \end{tabular}
    
    \vspace{0.4cm} 
    
    \begin{tabular}{cccc}
    \toprule
    \multicolumn{4}{c}{$\gamma=40.5$}\vspace{0.2cm}\\
    Time & Naive & Subtraction & Ratio  \\
    \midrule
    4 &  $3.71\times10^{-3}$ & $4.40\times10^{-5}$ & 84.2 \\
    6 & $4.81\times10^{-3}$ & $1.25\times10^{-4}$ & 38.5 \\
    8 & $6.36\times10^{-3}$ & $3.47\times10^{-4}$ & 18.3 \\ 
    \bottomrule   
    \end{tabular}
    \caption{Comparisons of variances between the naive transient and transient subtraction estimators for the Fourier coefficient $U_1$ in the transient dynamics~\eqref{eq:coupling} with friction coefficients of $\gamma=4.5$ and $\gamma=40.5$, corresponding to the results on the right panels in Figure~\ref{fig:transient equilibrium viscosity}.}
    \label{tab:coupling viscosity}
\end{table}

Table~\ref{tab:coupling viscosity} reports the variances of the estimations of $U_1$ at selected integration times, using the results from the right panels in Figure~\ref{fig:transient equilibrium viscosity}. Unlike the mobility case, the variance reduction from the transient subtraction estimator estimator is modest for $\gamma=4.5$. It can be seen from the values of the variance ratio for both cases that the transient subtraction estimator outperforms the naive transient estimator, especially for the higher friction limit, although the advantage of variance reduction gradually diminishes, as time increases. These results show that the relative variance reduction achieved by the transient subtraction estimator decreases as the integration time increases. Nevertheless, the transient subtraction estimator retains a lower variance than the naive transient estimator at all times considered.

\begin{figure}[tbp]
  \centering
  \includegraphics[scale=0.36]{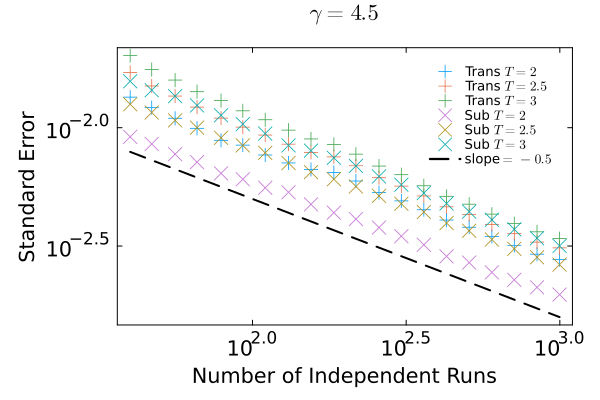}
  \includegraphics[scale=0.36]{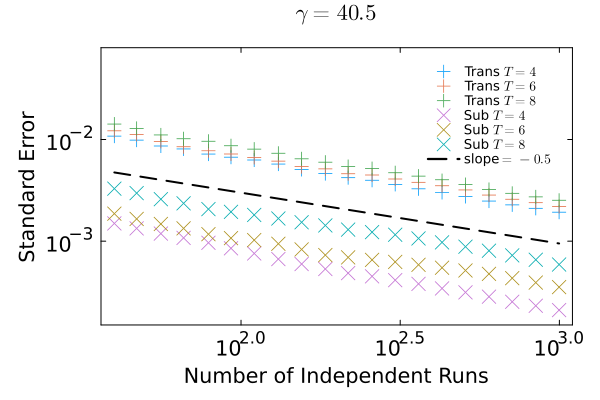}
  \caption{Standard errors of the response for the Fourier coefficient $U_1$~\eqref{eq:transient viscosity response} against the number of trajectories for the naive transient method and the transient subtraction method, with friction coefficients of $\gamma=4.5$ (left) and $\gamma=40.5$ (right) by using the sinusoidal force with the strength of $\eta=0.4$, in a log–log scale. The expected scaling line is plotted in dashed lines in all cases}
  \label{fig:se viscosity}
\end{figure}

The standard errors of the estimator $U_1$ as functions of the number of trajectories is plotted in Figure~\ref{fig:se viscosity}. It is observed that the standard errors decrease approximately in proportion to $N_{\mathrm{run}}^{-1/2}$ while their magnitudes depend on the estimator and the integration time for both friction coefficients. For $\gamma=4.5$, the standard error associated with the transient subtraction method is smaller than that of the naive transient method at each selected time, although the relative reduction becomes modest at later times. Meanwhile, increasing the number of trajectories reduces the standard errors of both estimators. A different behaviour is observed in the case of $\gamma=40.5$, where a clear separation between the two estimators persists at each selected time. Although this difference gradually diminishes at later times, the transient subtraction method maintains considerably smaller standard errors throughout the range of trajectories considered. These findings agree with the variance comparisons in Table~\ref{tab:coupling viscosity} and indicate that the transient subtraction method is more effective for estimating the Fourier response in the higher friction limit.

\section{Conclusions}
\label{sec:Conclusions}

In this article, we have studied the transient subtraction method for the computation of transport coefficients and investigated the variance reduction from the synchronous coupling in DPD. The method follows the relaxation of a system whose initial conditions are obtained by perturbing equilibrium samples. A reference trajectory, initialised at equilibrium, evolves under the same dynamics and Brownian noise. Its response serves as a control variate for the transient response. We have formulated this construction for the position-dependent noise in DPD and presented a finite-time variance analysis under stated regularity assumptions. The numerical experiments consider the mobility and the shear viscosity in a Lennard-Jones fluid. For shear viscosity, the transient subtraction estimator computes the Fourier response associated with the average longitudinal velocity, which is related to the viscosity through the formulation in~\cite{Wu2025}. These calculations have extended the numerical investigation of the transient subtraction method from Langevin dynamics~\cite{Spacek2025} to DPD.

The transient subtraction estimator requires the simulation of one equilibrium trajectory together
with each transient trajectory, and therefore increases the cost per run. Its efficiency should
therefore be assessed by comparing this additional cost with the variance reduction obtained
from the synchronous coupling. In the present experiments, the variance ratios are much
larger than this extra cost for most tested cases, especially in the high friction limit. This
indicates that the reduction in the sampling error is not merely a consequence of using more
trajectories, but comes from the correlation introduced by the synchronous coupling.

The numerical performance of the transient subtraction method exhibits slightly different trends in the computation of the mobility and the shear viscosity. For both transport coefficients, the transient subtraction method substantially reduces the variance relative to the naive transient dynamics in most of cases tested, particularly in the mobility computation, and the relaxation time is observed to be shorter than the start of the decoupling in all cases tested. It is worth mentioning that we have checked the cases of higher friction coefficients of $\gamma=200$ and $\gamma=450$ and observed more significant variance reduction as we increase $\gamma$, indicating that the performance of the transient subtraction method may be better in the higher friction limits. Moreover, we have verified that the standard errors decrease approximately as the inverse square root of the number of independent trajectories, consistent with the dependence on the number of trajectories in~\eqref{eq: var upper}. In the mobility computation, the estimations obtained from the naive transient dynamics exhibit larger fluctuations over the simulated intervals with friction coefficients of $\gamma=4.5$ and $\gamma=40.5$, while the transient subtraction method leads to much more accurate estimations for the transient dynamics comparing to the reference values obtained from the nonequilibrium approaches~\cite{Wu2025}. In the shear viscosity computation, even though the naive transient estimator approach approximately stationary values in both cases, the transient subtraction method produces more accurate estimations. Overall, the numerical results suggest that, the transient subtraction method improves the statistical efficiency in the computation of transport coefficients of the DPD system, and the benefit of employing the transient subtraction method becomes more pronounced as the friction coefficient increases. 

Furthermore, it will be of interest to extend the present analysis to more complex molecular dynamics models, including the single-particle DPD~\cite{Pan2008} and its extension to anisotropic particles~\cite{Deng2017}, where these models include rotational degrees of freedom. Future work could examine the performance of the transient subtraction method in computing their translational and rotational transport coefficients.

\section*{CRediT authorship contribution statement}

\textbf{Xinyi Wu:} Conceptualization, Methodology, Software, Validation, Formal analysis, Investigation, Writing - original Draft, Writing - review \& editing, Visualization. \textbf{Xiaocheng Shang:} Conceptualization, Methodology, Validation, Formal analysis, Investigation, Writing - original draft, Writing - review \& editing, Supervision, Funding acquisition.

\section*{Declaration of competing interest}

The authors declare that they have no known competing financial interests or personal relationships that could have appeared to influence the work reported in this paper.

\section*{Acknowledgements}

The authors thank Renato Spacek and Gabriel Stoltz for their valuable suggestions and comments. XW acknowledges the support from Research England under the Expanding Excellence in England (E3) funding stream which was awarded to MARS: Mathematics for AI in Real-world Systems in the School of Mathematical Sciences at Lancaster University, and the China Scholarship Council--University of Birmingham PhD Scholarship. XS acknowledges the support of the Royal Society through the International Exchanges Scheme (reference number IES$\backslash$R3$\backslash$203007), the Royal Society through the Research Grants (reference number RG$\backslash$R2$\backslash$232257), the Isaac Newton Institute for Mathematical Sciences through the Network Support for the Mathematical Sciences initiative (EPSRC reference number EP/V521929/1), the London Mathematical Society through the Emmy Noether Fellowship (reference number EN-2223-09), and CECAM and CCP5 via their CECAM/CCP5 sandpits. This work was supported by the Additional Funding Programme for Mathematical Sciences, delivered by (EPSRC EP/V521917/1 - Heilbronn Institute for Mathematical Research) and (EP/Z000580/1 - Isaac Newton Institute for Mathematical Sciences).

\appendix


\section{Proof of Proposition~\ref{prop1}}
\label{appendix:prop1}

Using the It\^{o}'s Lemma:
\begin{equation}
\begin{aligned}
  \dd \left( \left| \X_{t}^{\eta} - \Y_{t}^{0} \right|^{2} \right) = &\, 2 \left\langle \X_{t}^{\eta} - \Y_{t}^{0}, \bv\left(\X_{t}^{\eta}\right) - \bv\left(\Y_{t}^{0}\right) \right\rangle \, \dd t + \left\| \sigmaB\left(\X_{t}^{\eta}\right) - \sigmaB\left(\Y_{t}^{0}\right) \right\|^{2} \, \dd t \\
  &+ 2 \left\langle \X_{t}^{\eta} - \Y_{t}^{0}, \sigmaB\left(\X_{t}^{\eta}\right) - \sigmaB\left(\Y_{t}^{0}\right) \right\rangle \, \dd \WB_t \, . \label{eq:lemma last term}
\end{aligned}
\end{equation}
Taking expectations where the last term (the martingale) has mean zero:
\begin{equation}
  \frac{\dd}{\dd t} \E \left[ \left| \X_{t}^{\eta} - \Y_{t}^{0} \right|^{2} \right] \leqslant \left( 2B_{\bv} + B_{\sigmaB} \right) \E \left[ \left| \X_{t}^{\eta} - \Y_{t}^{0} \right|^{2} \right] \, .
\end{equation}
Then, we have
\begin{equation}
  \E \left[ \left| \X_{t}^{\eta} - \Y_{t}^{0} \right|^{2} \right] \leqslant e^{\left(2B_{\bv} + B_{\sigmaB} \right)t} \E \left[ \left| \X_{0}^{\eta} - \Y_{0}^{0} \right|^{2} \right] \, .
\end{equation}
Applying Gr\"{o}nwall's Lemma, we can derive~\eqref{eq:lemma}. Then, using the following Minkowski's integral inequality:
\begin{equation}
  \left( \E \left[ \left| \int^{T}_{0} \left| \X_{t}^{\eta} - \Y_{t}^{0} \right| \, \dd t \right|^{2} \right] \right)^{\frac{1}{2}} \leqslant \int^{T}_{0} \left( \E \left[ \left| \X_{t}^{\eta} - \Y_{t}^{0} \right|^{2} \right] \right)^{\frac{1}{2}} \, \dd t \, ,
\end{equation}
we have
\begin{subequations}
\begin{align}
  \Var \left( \widehat{\zeta}_{\eta,N_{\mathrm{run}},T}^{\mathrm{sub}} \right) & \leqslant \frac{1}{N_{\mathrm{run}}} \E \left[ \left| \widehat{\zeta}_{\eta,1,T}^{\mathrm{sub}} \right|^{2} \right] \\
  & \leqslant \frac{ \|R\|^{2}_{\mathrm{Lip}} }{ N_{\mathrm{run}}\eta^{2} } \E \left[ \left| \int^{T}_{0} \left| \X_{t}^{\eta} - \Y_{t}^{0} \right| \, \dd t \right|^{2} \right] \\
  & \leqslant \frac{ \|R\|^{2}_{\mathrm{Lip}} }{ N_{\mathrm{run}}\eta^{2} } \left\{ \int^{T}_{0} \left( \E \left[ \left| \X_{t}^{\eta} - \Y_{t}^{0} \right|^{2} \right] \right)^{\frac{1}{2}} \, \dd t \right\}^{2} \\
  & \leqslant \frac{ \|R\|^{2}_{\mathrm{Lip}} }{ N_{\mathrm{run}}\eta^{2} } \left\{ \int^{T}_{0} \left( e^{\left( 2B_{\bv} + B_{\sigmaB} \right)t} \E \left[ \left| \X_{0}^{\eta} - \Y_{0}^{0} \right|^{2} \right] \right)^{\frac{1}{2}} \, \dd t \right\}^{2} \\
  & = \frac{ \|R\|^{2}_{\mathrm{Lip}} }{ N_{\mathrm{run}}\eta^{2} } \E \left[ \left| \X_{0}^{\eta} - \Y_{0}^{0} \right|^{2} \right] \left( \int^{T}_{0} e^{\left( B_{\bv} + B_{\sigmaB}/2\right)t} \, \dd t \right)^{2} \, .
\end{align}
\end{subequations}



\bibliographystyle{elsarticle-num}


\bibliography{refs}

@Book{Allen2017,
  author    = {Allen, Michael P and Tildesley, Dominic J},
  publisher = {Oxford University Press, Oxford},
  title     = {Computer Simulation of Liquids, 2nd ed.},
  year      = {2017},
}

@Article{Arya2000,
  author    = {Arya, Gaurav and Maginn, Edward J. and Chang, Hsueh-Chia},
  title     = {Efficient viscosity estimation from molecular dynamics simulation via momentum impulse relaxation},
  journal   = {J. Chem. Phys.},
  year      = {2000},
  volume    = {113},
  number    = {6},
  pages     = {2079},
  issn      = {0021-9606},
  doi       = {10.1063/1.482019},
  publisher = {AIP Publishing},
  url       = {https://doi.org/10.1063/1.482019},
}

@Article{Backer2005,
  author    = {Backer, J. A. and Lowe, C. P. and Hoefsloot, H. C. J. and Iedema, P. D.},
  journal   = {J. Chem. Phys.},
  title     = {Poiseuille flow to measure the viscosity of particle model fluids},
  year      = {2005},
  pages     = {154503},
  volume    = {122},
  comment   = {A more accurate way to measure the viscosity in a particle model (including dissiparive particle dynamics, DPD) by using the Poiseuille flow method proposed in this article.

Excellent review of ways to calculate viscosity in particle methods.

Page 3: Conservative forces are not needed for this study and will be left out.},
  doi       = {10.1063/1.1883163},
  url       = {https://doi.org/10.1063/1.1883163},
}

@Article{Boromand2015,
  author    = {Boromand, Arman and Jamali, Safa and Maia, Joao M.},
  title     = {Viscosity measurement techniques in Dissipative Particle Dynamics},
  journal   = {Comput. Phys. Commun.},
  year      = {2015},
  volume    = {196},
  pages     = {149--160},
  issn      = {0010-4655},
  doi       = {10.1016/j.cpc.2015.05.027},
  publisher = {Elsevier BV},
  url       = {https://doi.org/10.1016/j.cpc.2015.05.027},
}

@Article{Chaudhri2010,
  author    = {Chaudhri, Anuj and Lukes, Jennifer R},
  journal   = {Phys. Rev. E},
  title     = {Velocity and stress autocorrelation decay in isothermal dissipative particle dynamics},
  year      = {2010},
  number    = {2},
  pages     = {026707},
  volume    = {81},
  comment   = {All numerical results are based on IDEAL DPD fluid.

Hi Xiaocheng,

Below the mentioned eq 29 they write that the system is isotropic, which is only the case in the absence of shear. So I think it must be the zero shear viscosity. On the other hand they call it somewhere a  dynamic shear viscosity, and this is again something else, the viscosity for the case of oscillatory shear at small amplitudes. This dynamic viscosity is complex-valued and can eventually be obtained via a Green-Kubo integral, but I am not 100% sure. In any case they messed it up. If you have no better evidence that the shear viscosity as function of rate can be obtained via some Green-Kubo integral, we better stay with the known relationship, which is an expression for the zero rate viscosity via Green-Kubo. Why do we mention Green-Kubo then? 

Martin},
  doi       = {10.1103/PhysRevE.81.026707},
  publisher = {APS},
  url       = {https://doi.org/10.1103/PhysRevE.81.026707},
}

@Article{Ciccotti1979,
  author    = {Ciccotti, G. and Jacucci, G. and McDonald, I. R.},
  title     = {``{T}hought-experiments'' by molecular dynamics},
  journal   = {J. Stat. Phys.},
  year      = {1979},
  volume    = {21},
  number    = {1},
  pages     = {1--22},
  issn      = {1572-9613},
  doi       = {10.1007/bf01011477},
  publisher = {Springer Science + Business Media},
  url       = {https://doi.org/10.1007/BF01011477},
}

@Article{Duenweg1993,
  author    = {D{\"u}nweg, Burkhard},
  title     = {Molecular dynamics algorithms and hydrodynamic screening},
  journal   = {J. Chem. Phys.},
  year      = {1993},
  volume    = {99},
  number    = {9},
  pages     = {6977--6982},
  doi       = {10.1063/1.465444},
  publisher = {AIP Publishing},
  url       = {https://doi.org/10.1063/1.465444},
}

@Article{Duncan2016,
  author    = {Duncan, A. B. and Leli{\`e}vre, T. and Pavliotis, G. A.},
  title     = {Variance Reduction Using Nonreversible {L}angevin Samplers},
  journal   = {J. Stat. Phys.},
  year      = {2016},
  volume    = {163},
  number    = {3},
  pages     = {457--491},
  issn      = {1572-9613},
  doi       = {10.1007/s10955-016-1491-2},
  publisher = {Springer Science + Business Media},
  url       = {https://doi.org/10.1007/s10955-016-1491-2},
}

@Article{Espanol1995,
  author    = {Espa{\~n}ol, Pep and Warren, Patrick},
  journal   = {Europhys. Lett.},
  title     = {Statistical mechanics of dissipative particle dynamics},
  year      = {1995},
  number    = {4},
  pages     = {191--196},
  volume    = {30},
  comment   = {Condition for weight function in the dissipative and the random contribution

for the DPD equations was derived, w_D(r)=w_R(r)^2, which was violated

in the original paper by Hoogerbrugge and Koelman (1992).},
  doi       = {10.1209/0295-5075/30/4/001},
  publisher = {IOP Publishing},
  url       = {https://doi.org/10.1209/0295-5075/30/4/001},
}

@Book{Evans2008,
  title     = {Statistical Mechanics of Nonequilibrium Liquids},
  publisher = {Cambridge University Press, Cambridge},
  year      = {2008},
  author    = {Evans, Denis J and Morriss, Gary},
}

@Book{Frenkel2001,
  author    = {Daan Frenkel and Berend Smit},
  publisher = {Academic Press},
  title     = {Understanding Molecular Simulation: From Algorithms to Applications, 2nd ed.},
  year      = {2001},
}

@Article{Green1954,
  author    = {Green, Melville S.},
  title     = {Markoff Random Processes and the Statistical Mechanics of Time-Dependent Phenomena. {II}. {I}rreversible Processes in Fluids},
  journal   = {J. Chem. Phys.},
  year      = {1954},
  volume    = {22},
  number    = {3},
  pages     = {398--413},
  doi       = {10.1063/1.1740082},
  url       = {https://doi.org/10.1063/1.1740082},
}

@Article{Hoogerbrugge1992,
  author    = {Hoogerbrugge, P. J. and Koelman, J. M. V. A.},
  title     = {Simulating microscopic hydrodynamic phenomena with dissipative particle dynamics},
  journal   = {Europhys. Lett.},
  year      = {1992},
  volume    = {19},
  number    = {3},
  pages     = {155--160},
  doi       = {10.1209/0295-5075/19/3/001},
  publisher = {IOP Publishing},
  url       = {https://doi.org/10.1209/0295-5075/19/3/001},
}

@Article{Kubo1957,
  author    = {Kubo, Ryogo},
  title     = {Statistical-Mechanical Theory of Irreversible Processes. {I}. {G}eneral Theory and Simple Applications to Magnetic and Conduction Problems},
  journal   = {J. Phys. Soc. Jpn.},
  year      = {1957},
  volume    = {12},
  number    = {6},
  pages     = {570--586},
  issn      = {1347-4073},
  doi       = {10.1143/jpsj.12.570},
  publisher = {Physical Society of Japan},
  url       = {https://doi.org/10.1143/jpsj.12.570},
}

@Book{Leimkuhler2015b,
  author    = {Leimkuhler, Ben and Matthews, Charles},
  publisher = {Springer, Cham, Switzerland},
  title     = {Molecular Dynamics: With Deterministic and Stochastic Numerical Methods},
  year      = {2015},
  comment   = {Modified Hamiltonian for the Verlet method is on Page 117 and 208.},
}

@Article{Leimkuhler2013c,
  author    = {Leimkuhler, B. and Matthews, C. and Stoltz, G.},
  journal   = {IMA J. Numer. Anal.},
  title     = {The computation of averages from equilibrium and nonequilibrium {L}angevin molecular dynamics},
  year      = {2016},
  issn      = {1464-3642},
  number    = {1},
  pages     = {13--79},
  volume    = {36},
  comment   = {Page 4: Nice discussion on why it is not always possible or desirable to use a Metropolis correction.},
  doi       = {10.1093/imanum/dru056},
  publisher = {Oxford University Press (OUP)},
  url       = {https://doi.org/10.1093/imanum/dru056},
}

@Article{Leimkuhler2015a,
  author    = {Leimkuhler, Benedict and Shang, Xiaocheng},
  title     = {Adaptive Thermostats for Noisy Gradient Systems},
  journal   = {SIAM J. Sci. Comput.},
  year      = {2016},
  volume    = {38},
  number    = {2},
  pages     = {A712--A736},
  issn      = {1095-7197},
  doi       = {10.1137/15M102318X},
  publisher = {Society for Industrial \& Applied Mathematics (SIAM)},
  url       = {https://doi.org/10.1137/15M102318X},
}

@Article{Leimkuhler2016a,
  author    = {Leimkuhler, Benedict and Shang, Xiaocheng},
  title     = {Pairwise adaptive thermostats for improved accuracy and stability in dissipative particle dynamics},
  journal   = {J. Comput. Phys.},
  year      = {2016},
  volume    = {324},
  pages     = {174--193},
  issn      = {0021-9991},
  doi       = {10.1016/j.jcp.2016.07.034},
  publisher = {Elsevier BV},
  url       = {https://doi.org/10.1016/j.jcp.2016.07.034},
}

@Article{Leimkuhler2015,
  author    = {Leimkuhler, Benedict and Shang, Xiaocheng},
  title     = {On the numerical treatment of dissipative particle dynamics and related systems},
  journal   = {J. Comput. Phys.},
  year      = {2015},
  volume    = {280},
  pages     = {72--95},
  issn      = {0021-9991},
  doi       = {10.1016/j.jcp.2014.09.008},
  publisher = {Elsevier BV},
  url       = {https://doi.org/10.1016/j.jcp.2014.09.008},
}

@Article{Pagonabarraga1998,
  author    = {Pagonabarraga, I. and Hagen, M. H. J. and Frenkel, D.},
  title     = {Self-consistent dissipative particle dynamics algorithm},
  journal   = {Europhys. Lett.},
  year      = {1998},
  volume    = {42},
  number    = {4},
  pages     = {377--382},
  doi       = {10.1209/epl/i1998-00258-6},
  publisher = {IOP Publishing},
  url       = {https://doi.org/10.1209/epl/i1998-00258-6},
}

@Article{Palmer1994,
  author    = {Palmer, Bruce J},
  journal   = {Phys. Rev. E},
  title     = {Transverse-current autocorrelation-function calculations of the shear viscosity for molecular liquids},
  year      = {1994},
  number    = {1},
  pages     = {359--366},
  volume    = {49},
  comment   = {Page 1: However, because there are only three unique off-diagonal components of J, the simulation must be run for long times to get highly converged values for the Green--Kubo integrand over the range of t values which contribute significantly to the integral.},
  doi       = {10.1103/physreve.49.359},
  publisher = {APS},
  url       = {https://doi.org/10.1103/physreve.49.359},
}

@Article{Pan2008,
  author    = {Pan, W. and Pivkin, I. V. and Karniadakis, G. E.},
  title     = {Single-particle hydrodynamics in {DPD}: A new formulation},
  journal   = {Europhys. Lett.},
  year      = {2008},
  volume    = {84},
  number    = {1},
  pages     = {10012},
  issn      = {1286-4854},
  doi       = {10.1209/0295-5075/84/10012},
  publisher = {IOP Publishing},
  url       = {https://doi.org/10.1209/0295-5075/84/10012},
}

@InProceedings{Shang2015,
  author    = {Shang, Xiaocheng and Zhu, Zhanxing and Leimkuhler, Benedict and Storkey, Amos J},
  booktitle = {Advances in Neural Information Processing Systems},
  title     = {Covariance-controlled adaptive {L}angevin thermostat for large-scale {B}ayesian sampling},
  year      = {2015},
  editor    = {C. Cortes and N. Lawrence and D. Lee and M. Sugiyama and R. Garnett},
  pages     = {37--45},
  publisher = {Curran Associates, Inc.},
  volume    = {28},
  url       = {https://papers.nips.cc/paper/5978-covariance-controlled-adaptive-langevin-thermostat-for-large-scale-bayesian-sampling},
}

@Article{Shardlow2006,
  author    = {Shardlow, Tony and Yan, Yubin},
  title     = {Geometric ergodicity for dissipative particle dynamics},
  journal   = {Stoch. Dyn.},
  year      = {2006},
  volume    = {6},
  number    = {01},
  pages     = {123--154},
  doi       = {10.1142/S0219493706001670},
  publisher = {World Scientific},
  url       = {https://doi.org/10.1142/S0219493706001670},
}

@Article{Verlet1967,
  author    = {Verlet, Loup},
  title     = {Computer ``experiments" on classical fluids. {I}. {T}hermodynamical properties of {L}ennard-{J}ones molecules},
  journal   = {Phys. Rev.},
  year      = {1967},
  volume    = {159},
  number    = {1},
  pages     = {98--103},
  doi       = {10.1103/PhysRev.159.98},
  publisher = {APS},
  url       = {https://doi.org/10.1103/PhysRev.159.98},
}

@Article{Wang2013a,
  author    = {Wang, Han and Hartmann, Carsten and Sch\"{u}tte, Christof},
  journal   = {Mol. Phys.},
  title     = {Linear response theory and optimal control for a molecular system under non-equilibrium conditions},
  year      = {2013},
  issn      = {1362-3028},
  number    = {22--23},
  pages     = {3555--3564},
  volume    = {111},
  doi       = {10.1080/00268976.2013.844370},
  publisher = {Informa UK Limited},
  url       = {https://doi.org/10.1080/00268976.2013.844370},
}

@Article{Shang2017,
  author    = {Shang, Xiaocheng and Kr{\"o}ger, Martin and Leimkuhler, Benedict},
  title     = {Assessing numerical methods for molecular and particle simulation},
  journal   = {Soft Matter},
  year      = {2017},
  volume    = {13},
  number    = {45},
  pages     = {8565--8578},
  doi       = {10.1039/C7SM01526G},
  publisher = {Royal Society of Chemistry},
  url       = {https://doi.org/10.1039/C7SM01526G},
}

@Article{Sachs2017,
  author    = {Sachs, Matthias and Leimkuhler, Benedict and Danos, Vincent},
  title     = {Langevin Dynamics with Variable Coefficients and Nonconservative Forces: From Stationary States to Numerical Methods},
  journal   = {Entropy},
  year      = {2017},
  volume    = {19},
  number    = {12},
  pages     = {647},
  doi       = {10.3390/e19120647},
  publisher = {Multidisciplinary Digital Publishing Institute},
  url       = {https://doi.org/10.3390/e19120647},
}

@Article{Espanol2017,
  author    = {Espa{\~n}ol, Pep and Warren, Patrick B},
  journal   = {J. Chem. Phys.},
  title     = {Perspective: {D}issipative particle dynamics},
  year      = {2017},
  number    = {15},
  pages     = {150901},
  volume    = {146},
  doi       = {10.1063/1.4979514},
  publisher = {AIP Publishing LLC},
  url       = {https://doi.org/10.1063/1.4979514},
}

@Article{Shang2019,
  author    = {Shang, Xiaocheng and Kr{\"o}ger, Martin},
  journal   = {SIAM Rev.},
  title     = {Time correlation functions of equilibrium and nonequilibrium {L}angevin dynamics: {D}erivations and numerics using random numbers},
  year      = {2020},
  number    = {4},
  pages     = {901--935},
  volume    = {62},
  doi       = {10.1137/19M1255471},
  url       = {https://doi.org/10.1137/19M1255471},
}

@Article{Lei2016a,
  author    = {Lei, Huan and Baker, Nathan A. and Li, Xiantao},
  journal   = {Proc. Natl. Acad. Sci. USA},
  title     = {Data-driven parameterization of the generalized {L}angevin equation},
  year      = {2016},
  number    = {50},
  pages     = {14183--14188},
  volume    = {113},
  comment   = {Supporting Information available: https://www.pnas.org/content/pnas/suppl/2016/11/29/1609587113.DCSupplemental/pnas.201609587SI.pdf},
  doi       = {10.1073/pnas.1609587113},
  publisher = {National Acad Sciences},
  url       = {https://doi.org/10.1073/pnas.1609587113},
}

@Article{Leimkuhler2019a,
  author    = {Leimkuhler, Benedict and Sachs, Matthias and Stoltz, Gabriel},
  title     = {Hypocoercivity properties of adaptive {L}angevin dynamics},
  journal   = {SIAM J. Appl. Math.},
  year      = {2020},
  volume    = {80},
  number    = {3},
  pages     = {1197--1222},
  doi       = {10.1137/19M1291649},
  publisher = {SIAM},
  url       = {https://doi.org/10.1137/19M1291649},
}

@Article{Roussel2019,
  author    = {Roussel, Julien and Stoltz, Gabriel},
  title     = {A perturbative approach to control variates in molecular dynamics},
  journal   = {Multiscale Model. Simul.},
  year      = {2019},
  volume    = {17},
  number    = {1},
  pages     = {552--591},
  doi       = {10.1137/18M1171047},
  publisher = {SIAM},
  url       = {https://doi.org/10.1137/18M1171047},
}

@Article{Shang2020,
  author    = {Shang, Xiaocheng},
  title     = {Accurate and efficient splitting methods for dissipative particle dynamics},
  journal   = {SIAM J. Sci. Comput.},
  year      = {2021},
  volume    = {43},
  number    = {3},
  pages     = {A1929--A1949},
  doi       = {10.1137/20M1336230},
  publisher = {SIAM},
  url       = {https://doi.org/10.1137/20M1336230},
}

@Article{Panoukidou2021,
  author    = {Panoukidou, Maria and Wand, Charlie R and Carbone, Paola},
  title     = {Comparison of equilibrium techniques for the viscosity calculation from {DPD} simulations},
  journal   = {Soft Matter},
  year      = {2021},
  volume    = {17},
  number    = {36},
  pages     = {8343--8353},
  doi       = {10.1039/D1SM00891A},
  publisher = {Royal Society of Chemistry},
  url       = {https://doi.org/10.1039/D1SM00891A},
}

@Article{Duong2021,
  author  = {Duong, Manh Hong and Shang, Xiaocheng},
  title   = {Accurate and robust splitting methods for the generalized {L}angevin equation with a positive {P}rony series memory kernel},
  journal = {J. Comput. Phys.},
  year    = {2022},
  volume  = {464},
  pages   = {111332},
  doi     = {10.1016/j.jcp.2022.111332},
  url     = {https://doi.org/10.1016/j.jcp.2022.111332},
}

@Article{Malaspina2023,
  author    = {Malaspina, D C and L{\'\i}sal, M and Larentzos, J P and Brennan, J K and Mackie, A D and Avalos, J Bonet},
  title     = {Transport coefficients from {E}instein--{H}elfand relations using standard and energy-conserving dissipative particle dynamics methods},
  journal   = {Phys. Chem. Chem. Phys.},
  year      = {2023},
  volume    = {25},
  number    = {17},
  pages     = {12025--12040},
  doi       = {10.1039/D2CP04838H},
  publisher = {Royal Society of Chemistry},
  url       = {https://doi.org/10.1039/D2CP04838H},
}

@Article{Pavliotis2022,
  author    = {Pavliotis, Grigorios A and Stoltz, Gabriel and Vaes, Urbain},
  title     = {Mobility estimation for {L}angevin dynamics using control variates},
  journal   = {Multiscale Modeling \& Simulation},
  year      = {2023},
  volume    = {21},
  number    = {2},
  pages     = {680--715},
  doi       = {10.1137/22M1504378},
  publisher = {SIAM},
  url       = {https://doi.org/10.1137/22M1504378},
}

@Article{Spacek2023,
  author    = {Spacek, Renato and Stoltz, Gabriel},
  title     = {Extending the regime of linear response with synthetic forcings},
  journal   = {Multiscale Model. Simul.},
  year      = {2023},
  volume    = {21},
  number    = {4},
  pages     = {1602--1643},
  doi       = {10.1137/23M1557611},
  publisher = {Society for Industrial \& Applied Mathematics (SIAM)},
  url       = {https://doi.org/10.1137/23M1557611},
}

@Article{Blassel2023,
  author    = {Blassel, No{\'e} and Stoltz, Gabriel},
  journal   = {J. Stat. Phys.},
  title     = {Fixing the flux: {A} dual approach to computing transport coefficients},
  year      = {2024},
  number    = {2},
  pages     = {17},
  volume    = {191},
  doi       = {10.1007/s10955-024-03230-x},
  publisher = {Springer},
  url       = {https://doi.org/10.1007/s10955-024-03230-x},
}

@Article{Malaspina2024,
  author    = {Malaspina, D. C. and Lísal, M. and Larentzos, J. P. and Brennan, J. K. and Mackie, A. D. and Avalos, J. Bonet},
  journal   = {Phys. Chem. Chem. Phys.},
  title     = {Green–{K}ubo expressions for transport coefficients from dissipative particle dynamics simulations revisited},
  year      = {2024},
  issn      = {1463-9084},
  number    = {2},
  pages     = {1328--1339},
  volume    = {26},
  doi       = {10.1039/D3CP03791F},
  publisher = {Royal Society of Chemistry (RSC)},
  url       = {https://doi.org/10.1039/D3CP03791F},
}

@Article{Greener2024,
  author    = {Greener, Joe G.},
  journal   = {Chem. Sci.},
  title     = {Differentiable simulation to develop molecular dynamics force fields for disordered proteins},
  year      = {2024},
  issn      = {2041-6539},
  number    = {13},
  pages     = {4897--4909},
  volume    = {15},
  doi       = {10.1039/D3SC05230C},
  publisher = {Royal Society of Chemistry (RSC)},
  url       = {https://doi.org/10.1039/D3SC05230C},
}

@InProceedings{Stoltz2022,
  author    = {Stoltz, Gabriel},
  booktitle = {Monte {C}arlo and {Q}uasi-{M}onte {C}arlo {M}ethods. {MCQMC} 2022.},
  title     = {Error Estimates and Variance Reduction for Nonequilibrium Stochastic Dynamics},
  year      = {2024},
  editor    = {Hinrichs, Aicke and Kritzer, Peter and Pillichshammer, Friedrich},
  pages     = {163--187},
  publisher = {Springer, Cham, Switzerland},
  series    = {Springer Proceedings in Mathematics \& Statistics},
  volume    = {460},
  doi       = {10.1007/978-3-031-59762-6_7},
  isbn      = {978-3-031-59762-6},
  url       = {https://doi.org/10.1007/978-3-031-59762-6_7},
}

@Article{Spacek2025,
  author    = {Spacek, Renato and Monmarché, Pierre and Stoltz, Gabriel},
  journal   = {J. Stat. Phys.},
  title     = {Transient Subtraction: A Control Variate Method for Computing Transport Coefficients},
  year      = {2025},
  issn      = {1572-9613},
  month     = apr,
  number    = {4},
  pages     = {53},
  volume    = {192},
  doi       = {10.1007/s10955-025-03424-x},
  publisher = {Springer Science and Business Media LLC},
  url       = {https://doi.org/10.1007/s10955-025-03424-x},
}

@Article{Melchior1995,
  author    = {Melchior, Martin and {\"O}ttinger, Hans Christian},
  journal   = {J. Chem. Phys.},
  title     = {Variance reduced simulations of stochastic differential equations},
  year      = {1995},
  issn      = {1089-7690},
  month     = dec,
  number    = {21},
  pages     = {9506--9509},
  volume    = {103},
  doi       = {10.1063/1.469961},
  publisher = {AIP Publishing},
  url       = {https://doi.org/10.1063/1.469961},
}

@Article{Wu2025,
  author    = {Wu, Xinyi and Shang, Xiaocheng},
  journal   = {J. Comput. Phys.},
  title     = {Stochastic {N}orton dynamics: {A}n alternative approach for the computation of transport coefficients in dissipative particle dynamics},
  year      = {2025},
  issn      = {0021-9991},
  month     = nov,
  pages     = {114316},
  volume    = {541},
  doi       = {10.1016/j.jcp.2025.114316},
  publisher = {Elsevier BV},
  url       = {https://doi.org/10.1016/j.jcp.2025.114316},
}

@Article{Lindvall1986,
  author    = {Lindvall, Torgny and Rogers, L. C. G.},
  journal   = {Ann. Probab.},
  title     = {Coupling of Multidimensional Diffusions by Reflection},
  year      = {1986},
  issn      = {0091-1798},
  month     = jul,
  number    = {3},
  pages     = {860--872},
  volume    = {14},
  doi       = {10.1214/aop/1176992442},
  publisher = {Institute of Mathematical Statistics},
  url       = {https://doi.org/10.1214/aop/1176992442},
}

@Article{Deng2017,
  author    = {Deng, Mingge and Pan, Wenxiao and Karniadakis, George Em},
  journal   = {J. Comput. Phys.},
  title     = {Anisotropic single-particle dissipative particle dynamics model},
  year      = {2017},
  issn      = {0021-9991},
  month     = may,
  pages     = {481--491},
  volume    = {336},
  doi       = {10.1016/j.jcp.2017.01.033},
  publisher = {Elsevier BV},
  url       = {https://doi.org/10.1016/j.jcp.2017.01.033},
}

@Article{Helfand1960,
  author    = {Helfand, Eugene},
  journal   = {Phys. Rev.},
  title     = {Transport Coefficients from Dissipation in a Canonical Ensemble},
  year      = {1960},
  issn      = {0031-899X},
  month     = July,
  number    = {1},
  pages     = {1--9},
  volume    = {119},
  doi       = {10.1103/PhysRev.119.1},
  publisher = {American Physical Society (APS)},
  url       = {https://doi.org/10.1103/PhysRev.119.1},
}

@Article{Espanol2009,
  author    = {Espa{\~n}ol, Pep},
  journal   = {Phys. Rev. E},
  title     = {Einstein--{H}elfand form for transport coefficients from coarse-grained descriptions},
  year      = {2009},
  issn      = {1550-2376},
  month     = Dec,
  number    = {6},
  pages     = {061113},
  volume    = {80},
  doi       = {10.1103/PhysRevE.80.061113},
  publisher = {American Physical Society (APS)},
  url       = {https://doi.org/10.1103/PhysRevE.80.061113},
}

@Article{Oettinger1994,
  author    = {{\"O}ttinger, Hans Christian},
  journal   = {Macromolecules},
  title     = {Variance Reduced {B}rownian Dynamics Simulations},
  year      = {1994},
  issn      = {1520-5835},
  month     = June,
  number    = {12},
  pages     = {3415--3423},
  volume    = {27},
  doi       = {10.1021/ma00090a041},
  publisher = {American Chemical Society (ACS)},
  url       = {https://doi.org/10.1021/ma00090a041},
}

@Article{Wagner1997,
  author    = {Wagner, Norman J. and {\"O}ttinger, Hans Christian},
  journal   = {J. Rheol.},
  title     = {Accurate simulation of linear viscoelastic properties by variance reduction through the use of control variates},
  year      = {1997},
  issn      = {1520-8516},
  month     = May,
  number    = {3},
  pages     = {757--768},
  volume    = {41},
  doi       = {10.1122/1.550831},
  publisher = {Society of Rheology},
  url       = {https://doi.org/10.1122/1.550831},
}

@Article{Eberle2019,
  author    = {Eberle, Andreas and Zimmer, Raphael},
  journal   = {Ann. Inst. H. Poincaré Probab. Statist.},
  title     = {Sticky couplings of multidimensional diffusions with different drifts},
  year      = {2019},
  issn      = {0246-0203},
  month     = Nov,
  number    = {4},
  pages     = {2370--2394},
  volume    = {55},
  doi       = {10.1214/18-AIHP951},
  publisher = {Institute of Mathematical Statistics},
  url       = {https://doi.org/10.1214/18-AIHP951},
}

@Article{Lampin2013,
  author    = {Lampin, E. and Palla, P. L. and Francioso, P.-A. and Cleri, F.},
  journal   = {J. Appl. Phys.},
  title     = {Thermal conductivity from approach-to-equilibrium molecular dynamics},
  year      = {2013},
  issn      = {1089-7550},
  month     = July,
  number    = {3},
  pages     = {033525},
  volume    = {114},
  doi       = {10.1063/1.4815945},
  publisher = {AIP Publishing},
  url       = {https://doi.org/10.1063/1.4815945},
}

@Article{Darshan2024,
  author  = {Shiva Darshan and Andreas Eberle and Gabriel Stoltz},
  journal = {arXiv:2409.15500},
  title   = {Sticky coupling as a control variate for sensitivity analysis},
  year    = {2024},
}

@Article{Blassel2026,
  author  = {No{\'{e}} Blassel and Louis Carillo and Shiva Darshan and Rapha{\"{e}}l Gastaldello and Alessandra Iacobucci and Elisa Marini and R{\'{e}}gis Santet and Xiaocheng Shang and Gabriel Stoltz and Urbain Vaes},
  journal = {arXiv:2605.10507},
  title   = {Mathematical analysis and numerical methods for the computation of transport coefficients in molecular dynamics},
  year    = {2026},
}

@Article{Ciccotti1975,
  author    = {Ciccotti, G. and Jacucci, G.},
  journal   = {Phys. Rev. Lett.},
  title     = {Direct Computation of Dynamical Response by Molecular Dynamics: The Mobility of a Charged {L}ennard-{J}ones Particle},
  year      = {1975},
  issn      = {0031-9007},
  month     = Sept,
  number    = {12},
  pages     = {789--792},
  volume    = {35},
  doi       = {10.1103/PhysRevLett.35.789},
  publisher = {American Physical Society (APS)},
  url       = {https://doi.org/10.1103/PhysRevLett.35.789},
}

\end{document}